\documentclass[11pt]{article}

\usepackage{fancyvrb}
\usepackage{color}
\usepackage{amsmath}
\usepackage{amsfonts}
\usepackage{amssymb}
\usepackage{amsthm}

\theoremstyle{definition}

\DefineVerbatimEnvironment{CodeInput}{Verbatim}{fontsize=\small,fontshape=sl,xleftmargin=2em}
\DefineVerbatimEnvironment{CodeOutput}{Verbatim}{fontsize=\small,xleftmargin=2em}

\DefineVerbatimEnvironment{Code}{Verbatim}{fontsize=\small,xleftmargin=2em,baselinestretch=1}
\usepackage[margin=1in]{geometry}
\usepackage{mathrsfs}
\usepackage{multirow}
\usepackage{booktabs}
\usepackage{threeparttable}
\usepackage{subcaption}
\usepackage{textcomp}
\usepackage{float}
\usepackage{hyperref}
\usepackage[numbers,sort&compress]{natbib}
\usepackage{graphicx}
\usepackage{authblk}
\usepackage{xcolor}

\usepackage{thmtools}

\usepackage{rotating}
\usepackage{array}
\newcolumntype{L}[1]{>{\raggedright\arraybackslash}p{#1}}

\title{A Bayesian hierarchical push-forward method for cluster Randomized Controlled Trial power calculations in complex humanitarian settings} 

\author[1]{Margaret Elliott{$*$}}
\author[1,2]{Sonia Jain, PhD}
\author[3,4]{Adam Richards, MD, PhD}
\author[5]{Brendan Fries}
\author[6]{Ravi Goyal, PhD}

\affil[1]{Division of Biostatistics, Herbert Wertheim School of Public Health and Human Longevity Science, University of California, San Diego USA}
\affil[2]{Biostatistics Research Center, University of California, San Diego, USA}
\affil[3]{Center for Humanitarian Health, Johns Hopkins Bloomberg School of Public Health, Maryland, USA}
\affil[4]{Community Partners International, Berkeley, California, USA}
\affil[5]{Milken Institute School of Public Health, The George Washington University, Washington D.C., USA}
\affil[6]{Division of Infectious Diseases and Global Public Health, University of California, San Diego USA}

\date{}

\begin{document}
\maketitle

\begin{center}
\small
$*$Corresponding author: meelliott@ucsd.edu\\
\end{center}

\begin{abstract}
The design and evaluation of public health interventions in humanitarian crises demand rigorous yet flexible methodological approaches. Cluster-randomized controlled trials (cRCTs) are increasingly used to generate evidence in such settings, but trial conduct is frequently complicated by displacement, resource constraints, and cluster dropout. Although Bayesian methods provide a natural framework for hierarchical modeling, probabilistic uncertainty quantification, and the integration of prior knowledge, their use in cRCT design remains limited, particularly with respect to anticipating missingness. This paper introduces a Bayesian design strategy tailored for cRCTs in complex settings, motivated by a forthcoming trial in Cox’s Bazar, Bangladesh, that will assess a community health worker adherence intervention for cardiovascular disease and hepatitis C. Our approach incorporates prior information on cluster and individual-level dropout into power and sample size calculations via a hierarchical model and push-forward distribution framework. Through simulation studies, we evaluate the performance of the proposed method under varying levels of missingness and correlation structures. The approach is then applied to inform the design of the Bangladesh trial. We disseminate our methods via a publicly-available R Shiny app to promote accessibility and usability to investigators. By formalizing Bayesian strategies that anticipate data loss, this work provides a foundation for more robust and efficient cRCTs in humanitarian contexts; therefore, strengthening the evidence base for health interventions in resource-limited and crisis-affected populations.
\end{abstract}


\section{Introduction}

Randomized clinical trials face many challenges to maintaining sufficient power to detect intervention effects in humanitarian settings, where time, personnel, and other resources are often extremely limited and triaged for emergent situations \citep{jpal2024,doocy2023,blanchet2015,burde2012}. Recent publications have called for development of new rigorous methods for randomized studies in complex settings to address challenges including low enrollment, delayed implementation of the intervention, contamination between study arms, poor intervention fidelity, and loss to follow-up \cite{blanchet2015, doocy2022}. In this paper, we develop a general Bayesian approach to mitigate the challenge of designing studies due to significant uncertainty in design features at both an individual- and cluster- level. Specifically, our novel approach estimates statistical power for a cRCT that incorporates uncertainty in key design characteristics, including intervention efficacy, intracluster correlation coefficients (ICC), and attrition. The approach was used to inform the design of a cluster-randomized controlled trial (cRCT) aimed at increasing cardiovascular disease (CVD) medication adherence and eliminating hepatitis C among refugees residing in Cox's Bazar, Bangladesh.

In the last two decades, cRCTs have been conducted in humanitarian settings to evaluate interventions for various chronic and acute health concerns \cite{doocy2023, doocy2022, blanchet2015}. Settings such as war, armed conflict, flooding, and political unrest have lead to internal and external displacement of people and dynamic logistical concerns. To conduct scientific studies in such settings, prespecified or ad hoc design considerations and adaptations are sometimes necessary, including the inclusion/exclusion of clusters, sample size considerations, stratified randomization, and flexible timelines and data collection methods \cite{watson2023, fenn2015, grijalva2023, leroy2021, tucker2024, jordans2021, kurdi2020, rahmaneffect2016, jordans2010, tol2014, kim2023, waltonclusterrandomized2025, fenn2017, roberts2001}. Information on cRCTs in complex settings and adaptations made is summarized in Appendix Table \ref{tab:tableA1}.

To proactively plan for trials in complex settings, it is necessary to understand the uncertainty in estimates of power, beyond a simple binary of whether the study is sufficiently powered or not. Researchers may consider the plausibility of a given effect size, and potential errors in the size or magnitude of estimates from previous studies. Alternative approaches to standard sample size calculations have been proposed that center the uncertainty in effect sizes. Gelman proposes new error types to quantify the uncertainty in the sign and magnitude of key parameter estimates, and Kunzmann advocates for a hybrid Bayesian-frequentist approach \cite{gelman2014, kunzmann2021}. These methods highlight the importance of thinking about power and new and nuanced ways, beyond the traditional frequentist framework. 

The Bayesian paradigm offers an appropriate and popular choice to estimate uncertainty as well as allow for hierarchical modeling and the incorporation of prior information. However, a 2021 systematic review found a limited number of Bayesian applications to cRCTs in the published literature \cite{jones2021}. In particular, there is a lack of Bayesian methods developed for the design stage of cRCTs. Existing design approaches include using posterior distributions of ICC for sample size and power calculations \cite{turner2004, turner2005, spiegelhalter2001}, sample size determination via cost-benefit analysis within the pharmaceutical development pipeline \cite{kikuchi2010} and sample size re-estimation via an internal pilot study \cite{lakesample2002}. Stepped wedge cRCTs (SWCRTs) are another design approach often considered when logistical or ethical constraints are present, or when temporal trends are of interest \cite{grayling2017}. Wang et al. propose a Bayesian adaptive stepped wedge design for early stopping determined by interim efficacy/futility analysis \cite{wang2024}. Kasza et al. propose a batched SWCRT to combat challenges introduced by recruitment delays \cite{kasza2022}, and efficient incomplete SWCRT designs have been considered \cite{rezaeidarzi2025}. Despite these existing methods, there remains a marked lack of strong health intervention evidence and adaptive design approaches for cluster randomized trials in humanitarian crisis settings \cite{blanchetevidence2017, doocy2023}. There is especially a need for statistical design methods focused on missingness in cRCTs due to complex emergency settings. 

This paper is motivated by a cluster randomized trial in Bangladesh that will evaluate a screening and treatment intervention for cardiovascular disease (CVD). In Cox's Bazar, Bangladesh there is a large refugee camp of displaced people from Myanmar. CVD is highly prevalent in this population and is a significant health burden \cite{rahman2022, schramm2025}. The cRCT plans to assess a combined adherence intervention administered by community healthcare workers for treatment of CVD in the refugee population \cite{thwe2026}.

A key concern for the Bangladesh study, as well as other cRCT studies in humanitarian settings, is study attrition. Furthermore, the follow-up rates may be heterogeneous across clusters, and there is the possibility that entire clusters may be lost. Various approaches exist to account for missingness at the analysis stage, including Bayesian methods \cite{rubin2018}. The method presented aims to account for potential missing data at the design stage by developing a Bayesian framework that informs power and sample size calculations with prior information related to key design considerations. Specifically, we propose a Bayesian design framework that uses data from prior trials to jointly model intervention efficacy and attrition in cluster randomized settings. Posterior uncertainty in key parameters, including treatment effect, follow-up rate, and ICC, is propagated through sample size formulas to obtain a ``push-forward'' predictive distribution for power. We further extend this framework using hierarchical models to borrow information across multiple related studies. 

The remainder of the paper is structured as follows. The next section summarizes existing methods that are relevant to our approach. Section~\ref{sec:methods} describes the proposed approach, consisting of a hierarchical model and a push-forward distribution for power. Section~\ref{sec:sim_study} presents a simulation study to illustrate the proposed methods by estimating the push-forward power distribution under a range of parameter values. Section~\ref{sec:application} presents additional details and power results for our primary application of designing a cRCT in Bangladesh. In Section~\ref{sec:discussion}, we describe our conclusions, note limitations, highlight promising areas for future research, and discuss the R Shiny application available for practitioners to use to design their studies.

\section{Background}

Our framework integrates two areas of methodological work to aid the design of trials in complex settings. First, we extend a Bayesian push-forward approach that characterizes uncertainty in the ICC and propagates this uncertainty into a predictive distribution for trial power \cite{turner2004, turner2005}. Second, we incorporate established sample size formulations that account for attrition at both the individual- and cluster- level through variance inflation factors \cite{donner2000}. In complex settings, where follow-up and cluster retention are highly uncertain, combining these approaches allows uncertainty in both outcome correlation and missingness to be formally reflected in design-stage power assessments.

\subsection{Bayesian Push-forward Distribution Framework} \label{sec:background_PFPD}

The ICC provides a metric summarizing the magnitude that measurements within a cluster are correlated and is a foundational parameter in designing cRCTs. In conventional practice, a point estimate of the ICC is used for sample size determination, yielding a single point estimate of the power and corresponding confidence interval; see \cite{donner2000} for details on sample size formulas for cRCTs. However, ICC estimates at the design stage are often imprecise, such uncertainty is not represented by a point estimate.

To quantify beliefs about the ICC and incorporate uncertainty in the estimated power of a planned trial, Turner et al. present Bayesian approaches to ICC estimation \cite{turner2004, turner2005}. Using an existing ICC estimate from a completed trial, they derive a posterior distribution of the ICC for a planned trial, which may be further used to construct a distribution for power \cite{turner2004}. Furthermore, they present hierarchical models to relate ICC estimates from multiple studies, and allow for the incorporation of outcomes data \cite{turner2005}.  We refer to this process of using posterior distributions for study design parameters to construct a distribution for power as creating a "push-forward" power distribution. Such methods allow for prior information, potentially from multiples sources, to be used in creating a distributional representation of trial operating characteristics, which provides more nuanced information as compared to a point estimate.

\subsection{Sample Size Calculations for cRCTs with Attrition} \label{sec:background_sample_size}

Several methods for cluster-randomized trial sample size calculations are available that account for loss-to-follow-up. Complexities due to attrition are typically accounted for with a design effect or variance inflation factor \cite{rutterford2015}. We focus on sample size formulas with explicit parameters for attrition due to our interest in designing cRCTs in complex humanitarian settings. Specifically, we use the following sample size formula for binary outcomes, with design effect $G$ to account for attrition, as described by Taljaard et al.\cite{taljaard2007}:

\begin{align}
    n \times m &= \frac{(z_{\alpha / 2} + z_{\beta})^2 [P_1(1-P_1) + P_0(1-P_0)]}{\delta_c^2} \times \frac{2G}{\pi_c}\\
 \notag G &= 1 + (m\pi_c - 1)\rho_c + (1-\pi_c)(1 + (m-1)\tau_c)\rho_c.
\end{align}

The intraclass correlation coefficient for attrition, $\tau_c$, is defined the same as the outcome ICC $\rho_c$, only it is in terms of attrition instead of the outcome. The quantity describes how much correlation between two individuals from the same cluster having missing observations. At small values near 0, individuals in a cluster are not or only slightly correlated, and missingness is nearly random and not dependent on cluster residence. At values near 1, individuals in a cluster are highly correlated; this captures when entire clusters are lost to follow-up. 

\noindent Parameter descriptions and notations are summarized in Table~\ref{tab:sample_size_formula_notation}. Other sample size formulae with parameters for attrition can be used without modification of the theory presented. 

\begin{table}[ht!]
    \centering
    \begin{tabular}{|c|l|}
        \hline
        \textbf{Parameter} & \textbf{Definition} \\\hline
         $n$ & total number of clusters \\
         $m$ & number of individuals per cluster \\
         $\alpha$ & probability of Type I error \\
         $1 - \beta$ & power \\
         $P_1$ & outcome proportion in experimental arm  \\
         $P_0$ & outcome proportion in control arm\\
         $\delta_c$ & treatment effect size, equal to $P_1 - P_0$ \\
         $\pi_c$ & follow-up rate for setting $c$ \\
         $\rho_c$ & ICC for outcome for setting $c$\\
         $\tau_c$ & ICC for attrition for setting $c$ \\\hline
    \end{tabular}
    \caption{Statistical notation used for a cluster-randomized trial with attrition and a binary outcome.}
    \label{tab:sample_size_formula_notation}
\end{table}
 
\section{Methods} \label{sec:methods}

In this section, we present a Bayesian framework for estimating a push-forward power distribution (PFPD), denoted as $P(\xi \mid D, \Gamma)$, for a future cluster randomized trial. This framework incorporates information and uncertainty about key design parameters, including attrition, into the PFPD. 

To evaluate the power of a new trial, investigators specify a set of design constraints, $\Gamma$, which can consist of the planned number of clusters $n$, the number of individuals per cluster $m$, and the desired type I $(\alpha)$ and type II $(\beta)$ error rates; therefore, $\Gamma = \{n, m, \alpha, \beta\}$. Investigators must also specify values for certain parameters $\Theta$, depending on the formula being used. In our setting, $\Theta$ consists of the primary sources of trial uncertainty: the treatment effect $\delta_c$, the outcome ICC $\rho_c$, and parameters governing individual- and cluster-level attrition $\pi_c$ and $\tau_c$, respectively. Rather than relying on point estimates for $\Theta$, our approach utilizes historical data, $D$, from one or more previously completed studies and other relevant prior information to estimate a posterior distribution for $\Theta$. We may then draw posterior samples $\Theta_{new}$ of parameter values for the planned trial.

The PFPD is constructed by defining a deterministic mapping function, $f: (\Theta_{\text{new}}, \Gamma) \mapsto \xi$, which links the fixed design constraints to key design parameters, $\Theta_{\text{new}} \in \mathbb{R}^p$, where $p$ is the dimension of the design parameters. By propagating uncertainty codified by the posterior distribution for $\Theta$ through the mapping function $f$, we induce the PFPD. The following subsections detail the mathematical formulation of this push-forward mapping function $f$ (Section~\ref{sec:PFPD}), the hierarchical models used to derive the posterior distributions from single or multiple prior trials (Section~\ref{sec:post_distr}), and the specific probability models applied to attrition and intervention effects.

\subsection{Push-forward Power Distribution} \label{sec:PFPD}

Given $D$ and $\Gamma$, we may write the PFPD as:

\begin{equation} \label{eq:PFPDeqn}
\begin{split}
& P(\xi \mid D, \Gamma) = \int_{\Phi} \int_{\Theta_{\text{new}}} \mathbb{I}\{f(\Theta_{\text{new}}, \Gamma)=\xi\}\, p(\Theta_{\text{new}} \mid \Phi)\,  p(\Phi \mid D)\,d\Theta_{\text{new}} d\Phi,
\end{split}
\end{equation}

\noindent where $p(\Phi \mid D)$ is the marginal posterior distribution of the hyperparameters given past data, and $p(\Theta_{\text{new}} \mid \Phi)$ represents the population distribution from which the future trial's parameters are drawn.

There is not a closed-form solution for the integral; therefore, it is approximated using Monte Carlo integration. Specifically, we draw $\{\Phi^{(b)}\}_{b=1}^B$ from the posterior distribution $p(\Phi \mid D)$, sample a new candidate parameter vector $\Theta^{(b)}_{\text{new}} \sim p(\cdot \mid \Phi^{(b)})$, and compute the corresponding power values $\xi^{(b)} = f(\Theta^{(b)}_{\text{new}}, \Gamma)$. The resulting empirical PFPD, $\{\xi^{(b)}\}_{b=1}^B$,  provides a numerical approximation to $P(\xi \mid D, \Gamma)$; this is an extension and formalization of Turner et al.'s framework for calculating power based on a posterior distribution for the ICC \cite{turner2004}. From the empirical PFPD, summary measures such as the mean, median, credible intervals, and tail probabilities can be obtained. This distribution characterizes uncertainty in trial power induced by uncertainty in the underlying design parameters.

\subsection{Posterior Distribution} \label{sec:post_distr}

To incorporate information from multiple trials into the power distribution, we define a hierarchical model relating trial-specific parameters across multiple completed studies. Let $c \in \{1, 2,\ldots, C\}$ denote a given trial or setting with parameters $\Theta_c$. We assume these trial-specific parameters are independently and identically distributed around hyperparameters $\Phi$. Let $\boldsymbol{\Theta} = \{\Theta_1, \dots, \Theta_C\}$ represent the collection of all historical trial parameters. 

The joint posterior $p(\boldsymbol{\Theta}, \Phi | D)$ on the hyperparameters and parameters is:

\begin{equation}
    p(\boldsymbol{\Theta}, \Phi \mid D) \propto L(D \mid \boldsymbol{\Theta}) \times p_0(\boldsymbol{\Theta}, \Phi),
\end{equation}

\noindent where $D$ denotes all the available data from the multiple trials and the prior distribution $p_0(\boldsymbol{\Theta}, \Phi)$ is the following:

\begin{equation}
   p_0(\boldsymbol{\Theta}, \Phi) =  p_0(\Phi) \times \prod_{c=1}^C p(\Theta_c \mid \Phi).
\end{equation}

\noindent Therefore,

\begin{equation}
 p(\boldsymbol{\Theta}, \Phi \mid D) \propto L(D \mid \boldsymbol{\Theta}, \Phi) \times p_0(\Phi) \times \prod_{c=1}^C p(\Theta_c \mid \Phi).
\end{equation}

The data from trial $c$ is denoted as $D_c = (Y_c, R_c)$,  where $Y_c$ and $R_c$ are the outcome and attrition data, respectively. Assuming all $D_c$ are independent given $\Phi$ and the outcome is independent of the attrition data, we have:

\begin{align}
p(\boldsymbol{\Theta}, \Phi \mid D) &\propto p_0(\Phi) \times \prod_c L(D_c | \Theta_c)p(\Theta_c|\Phi)\\\notag
&\propto p_0(\Phi) \times \prod_c L(Y_c | \Theta_c)L(R_c|\Theta_c)p(\Theta_c|\Phi).
\end{align}

In the following section we provide details for $L(Y_c | \Theta_c)$ and $L(R_c|\Theta_c)$ motivated by the application in Section~\ref{sec:application}. In our simulation study (Section~\ref{sec:sim_study}) and application (Section~\ref{sec:application}), we use independent uniform priors for all elements of $\Theta_c$.

\subsection{Attrition and Outcome Models}

The proposed approach is general and can accommodate a range of statistical formulations for attrition and outcome models. In this section, we specify the precise models utilized in the simulation study (Section~\ref{sec:sim_study}) and primary application (Section~\ref{sec:application}). The PFPD approach can use individual-level data, or take parameter estimates and simulate individual-level data; the latter is implemented in the Shiny application.  

For a given trial $c$, we are interested in estimating the follow-up rate $\pi_c$, the attrition ICC $\tau_c$, the outcome ICC $\rho_c$ (as detailed in Section~\ref{sec:background_sample_size}) as well as the effect parameters $\beta_{0c}$ and $\beta_{1c}$ (described below) from a single trial. We denote these parameters as a trial-specific vector $\Theta_c = (\pi_c, \tau_c, \beta_{0c}, \beta_{1c}, \rho_c)$.

\subsubsection{Attrition Model} \label{sec:attribution_model}

Let $R_{ikc} = 1$ denote if the follow-up outcome measurement for individual $i$ in cluster $k$ from trial $c$ is observed. We assume that within a cluster, all individuals have the same propensity of their outcome being observed, denoted as $R_{kc}$; however, the propensity may vary across cluster and trial. We assume a missing completely at random attrition structure, but missingness could be conditioned on covariates. The proportion of non-missing follow-up observations in each cluster $i \in \{1,...,n\}$ of size $m_i$ in trial $c$ are assumed to be Beta distributed random variables; therefore, $R_{ic} \sim Beta(a_c, b_c)$.

This allows us to express the mean follow-up rate $\pi_c$ and the attrition ICC $\tau_c$ for trial $c$ in terms of the mean and variance of the Beta distribution as:

\begin{align}
\notag \pi_c &:= E(R_{kc})  = \frac{a_c}{a_c + b_c} \\ 
\tau_c &:= \frac{Var(R_{kc})}{Var(R_{kc})} =  \frac{\sigma^2_{R_{kc}}}{\pi_c(1-\pi_c)} = \frac{1}{a_c + b_c + 1}.
\end{align}

This formulation assumes that all clusters share a common mean follow-up rate \cite{eldridge2009}. While a logistic model for the individual attrition indicators could alternatively be specified, it lacks a closed-form expression for the ICC ($\tau$), requiring the value to be approximated.  Full details regarding the reparameterization of $\tau$ in terms of $a$ and $b$ are available in Appendix A.2. 

By treating the cluster-specific propensities $R_{kc}$ as latent variables and integrating them out, we define the likelihood of the attrition data $R_c$ given the trial-specific parameters $\Theta_c$ as:

\begin{equation}
L(R_c \mid \Theta_c) = \prod_k \int_0^1 f_B(R_{kc} \mid \pi_c, \tau_c) \bigg[\prod_i f_{Bn}(R_{ikc} \mid R_{kc})\bigg] \, dR_{kc},
\end{equation}

\noindent where $f_B$ and $f_{Bn}$ are the density functions for the Beta and Bernoulli distributions. 

\subsubsection{Outcome Model} \label{sec:outcome_model}

Let $Y_{ikc}$ denote the outcome of individual $i$ in cluster $k$ in trial $c$ and let $X_{ikc}$ denote a vector of individual or cluster-level covariates, including arm assignment. In our application of interest, the primary outcome of treatment adherence is binary, therefore, we use a generalized linear mixed effects model:

\begin{align} \label{eq:outcome_model}
\notag &Y_{ikc} \sim Bern(p_{ikc})\\ 
&logit(p_{ikc}) =  \beta^TX_{ikc} + \eta_{kc},
\end{align}

\noindent where we assume the cluster-level ($\eta_{ic}$) random effect is normally distributed with mean 0 and variances $\sigma^2_{\eta_c}$.  The outcome ICC for trial $c$ can be approximated as \cite{eldridge2009}:

\begin{align}
    \rho_c \approx \frac{\sigma^2_{\eta_c}}{\sigma^2_{\eta_c} + (\pi^2/3)}, 
\end{align}

where $\pi$ is the numeric quantity $\approx 3.14$. 

The intervention effect size $\delta_c$ is defined as the difference in population adherence rate between arms $\delta_c := P_{1c} - P_{0c}$ where $P_{0c}$ and $P_{1c}$ denote the outcome probability in the control and intervention arms, respectively, for trial $c$. Although, we use a binary outcome model as it is relevant to the application presented in Section 5, but these methods could be extended to a continuous outcome or to multiple outcomes. 

By treating $\eta_{kc}$ as a latent variable and integrating, we have the following for the likelihood for the outcome of trial $c$:

\begin{align}
\notag L(Y_c \mid \Theta_c) = &\prod_{k=1}^n \int_{-\infty}^{\infty} \left[ \prod_{i=1}^{m} f_{\text{Bn}}\left(Y_{ikc} \mid \text{logit}^{-1}(\beta_{0c} + \beta_{1c}X_{ikc} + \eta_{kc})\right) \right] \\
                    & \times f_{\text{N}}\left(\eta_{kc} \mid 0, \frac{\rho_c}{1 - \rho_c}\frac{\pi^2}{3}\right) \, d\eta_{kc}
\end{align}

\noindent where $f_{Bn}$ and $f_N$ are the density functions for the Bernoulli and normal distributions, respectively. 

\section{Simulation Study} \label{sec:sim_study}

We conducted a simulation study to investigate the performance and behavior of the presented method to estimate power for a new cRCT across different settings. The simulations investigate varying sample sizes and varying outcome parameters. We use a trial design similar to that of the Bangladesh application presented in Section \ref{sec:application}, where the primary outcome is adherence to medication. We first outline the data generation process for producing simulated trial datasets (Section~\ref{sec:sim_data}) that represent completed but similar trials as the potential study, and then present results for the simulations on estimated power.

\subsection{Synthetic Data Generation} \label{sec:sim_data}

In this section, we provide details on generating synthetic data for a single cRCT $c$ with a binary outcome; as noted above, this data represents a previously concluded cRCT similar to the proposed study. The binary outcome aligns with the proposed Bangladesh study (See Section~\ref{sec:application}) investigating CVD adherence.

The probability of the outcome $p_{ikc}$ for each synthetic individual is generated based on the model in Equation~\ref{eq:outcome_model} from Section~\ref{sec:outcome_model}, with $X_{ikc}$ indicating whether cluster $k$ is assigned to the experimental treatment. We assume a 1:1 allocation ratio, that is, the same number of clusters are in each arm. Adherence outcomes ($Y_{ikc}$) are then generated as Bernoulli random variables with probability $p_{ikc}$. Motivated by the Bangladesh cRCT, we assume that the adherence proportion in the control clusters is 30\%, corresponding to $\beta_0 = -0.85$, and use $\sigma_{\eta_{kc}} = 0.05$. We vary the treatment effect ($\beta_1$) across the simulation study, corresponding to $\delta$ ranging from 0.10 to 0.70 by increments of 0.05.

To simulate attrition, we generate the follow-up proportion $R_{kc}$ in each cluster $k$ as:

\begin{align}
R_{kc} &\sim Beta(a_c, b_c) \\
\notag a_c &= \frac{\pi_c(1-\tau_c)}{\tau_c} \\
\notag b_c &= \frac{1}{\tau_c}-1-a_c,
\end{align}

\noindent where the overall follow-up rate is $\pi_c$ and cluster attrition heterogeneity of $\tau_c$ for trial $c$. We then simulate follow-up indicator variables for each individual $k$ in cluster $i$ as Bernoulli random variables with probability $w_i$. For the simulation study, we vary $\pi_c$ and $\tau_c$ from 0.10 to 0.95 by increments of 0.05.

We simulate data and construct a push-forward distribution for all combinations of the follow-up rate, attrition ICC, and treatment effect, with $m = 20$, $n = 10$, $\alpha = 0.05$, $P_1 = 0.30$, and $\sigma_{\eta} = 0.05$. For each data set, $5$ replicates are generated. To calculate frequentist power values for comparison, we take the parameters used to generate the simulated data set. 

\subsection{Simulation Results} \label{sec:simulation}

For each simulated dataset $D$, we applied the Bayesian estimation framework and constructed the corresponding PFPD using the probability models described in Sections~\ref{sec:attribution_model} and ~\ref{sec:outcome_model}. Our simulation study to evaluate the characteristics of the PFPDs consists of two investigations. First, we examine the distribution under various cluster sizes ($m$) and number of clusters ($n$). Second, we investigate the distribution for various trial treatment effects ($\delta_c$).

\subsubsection{Sample Size} \label{sec:sim_samplesize}

\begin{figure*}[ht!]
    \centering
    \includegraphics[width=6in]{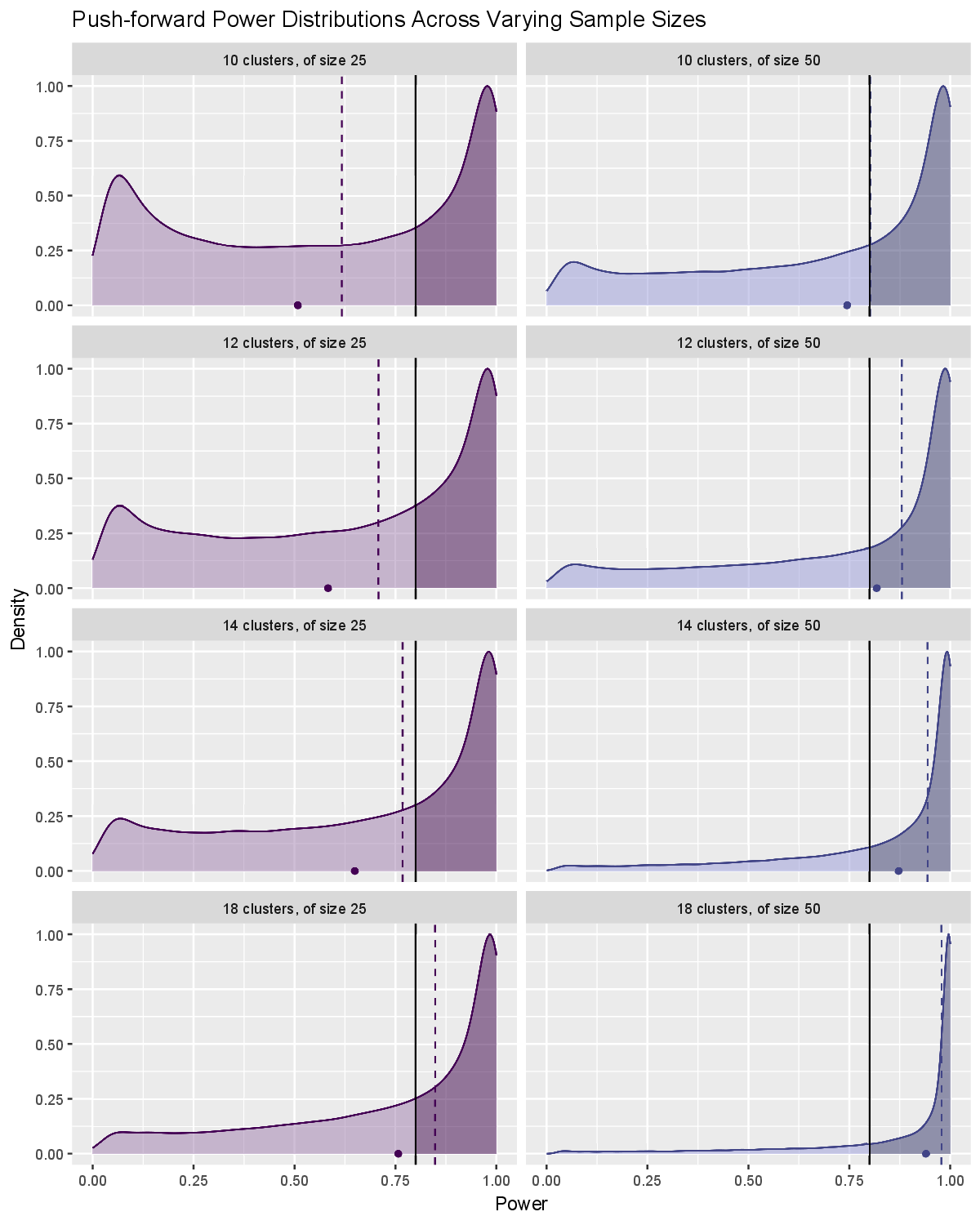}
    \caption{Push forward power distributions across varying sample sizes. In each panel of Figure \ref{fig:figure1}, the solid line is at $80\%$ power; the dashed line is the median power based from the PFPD; and the dot is the frequentist point estimate. For a given number of clusters, each plot shows the PFPD as the cluster size increases (left to right). Each distribution curve is the result of 50 replicates of the given parameter combination, with a follow-up rate of 42\%, an effect size of 20\%, an ICC for attrition of 0.06, a 0.30 outcome proportion in the control arm, a type I error rate of 0.05, and two experimental arms.}
    \label{fig:figure1}
\end{figure*}

We present the push-forward power distributions for varying sample sizes in Figure~\ref{fig:figure1}. Each row in the figure corresponds to a fixed number of clusters ($n \in \{10, 12, 14, 18\}$), while the columns correspond to a cluster size ($m \in \{25, 50\}$). Each distribution curve is the result of 50 replicates of the given parameter combination, with a follow-up rate of 42\%, an effect size of 20\%, an ICC for attrition of 0.06, a 0.30 outcome proportion in the control arm, a type I error rate of 0.05, and two experimental arms. As expected, the power density distributions shift to the right and become concentrated at higher power values as the number of clusters or individuals per cluster increases. At lower cluster sizes (see the left-most column of Figure \ref{fig:figure1}), the peak near 0 is the result of the treatment effect posterior distribution having density at negative values. This result may be seen consistently at middling treatment effect sizes (see the center left panels of Figures \ref{fig:figure1} and \ref{fig:figure2}), where the continuous treatment effect distribution has a significant amount of density below 0. 

In each panel of Figure \ref{fig:figure1}, the solid line is at $80\%$ power; the dashed line is the median power based from the PFPD; and the dot is the frequentist point estimate. In the top right panel of Figure \ref{fig:figure1}, for 10 clusters of size 50, the median of the PFPD is exactly at 80\% power (the solid and dashed line overlap). While this summary statistic may initially convey that the trial is sufficiently powered, a more thorough examination of the PFPD tells a different story: There is a 50\% chance that the trial is not sufficiently powered at 80\%, and 28\% of the power density is below 50\%. This indicates that the real trial may be considerably underpowered, which has ramifications for estimating spurious associations \cite{gelman2014}. This general pattern can be see in several of the panels (e.g., 12 clusters of size 50). 

\subsubsection{Attrition and Treatment Effect} \label{sec:attrition_adherence}

\begin{figure*}[ht!]
    \centering
    \includegraphics[width=6in]{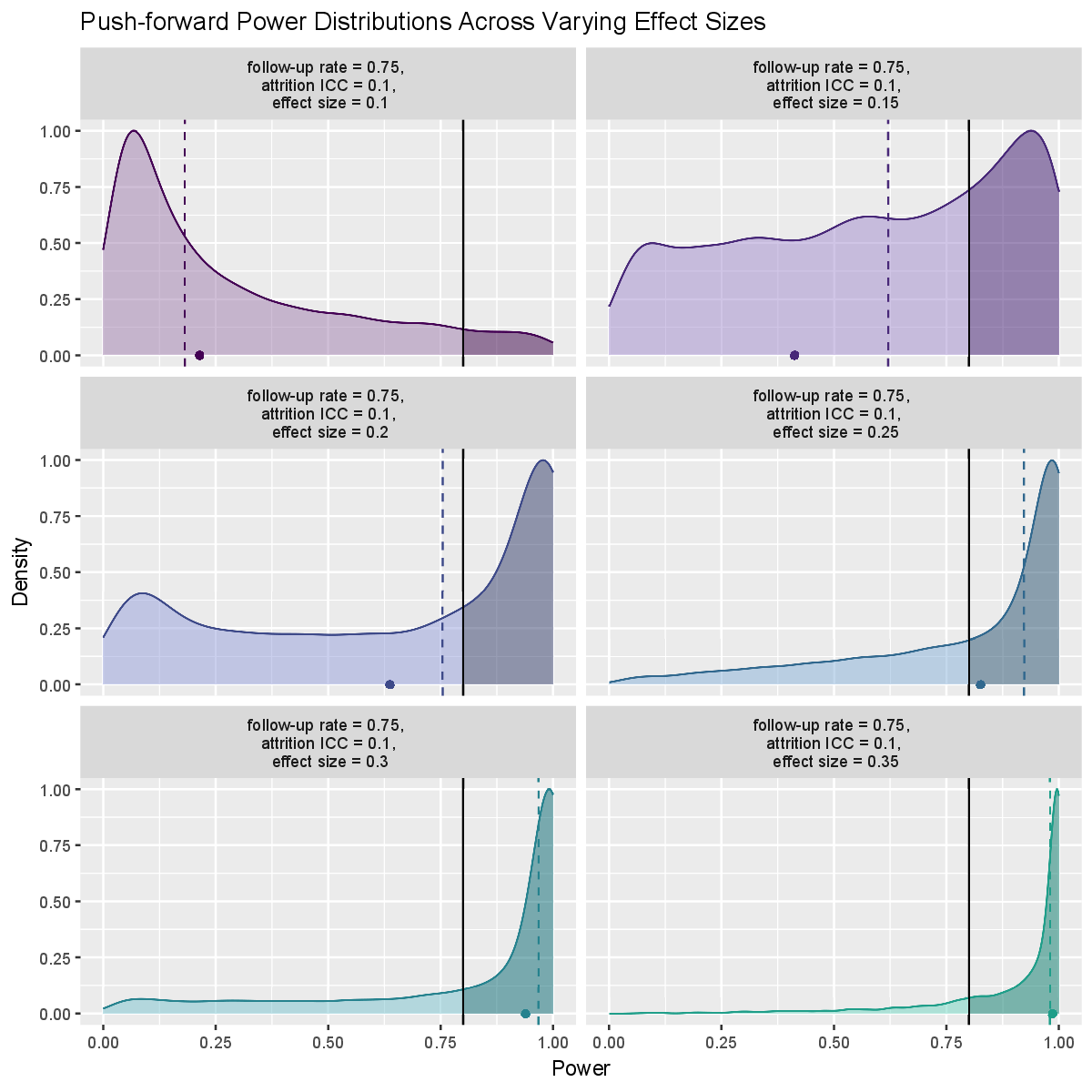}
    \caption{Push forward power distributions across varying treatment effect sizes (0.10, 0.20, 0.30, 0.40) given a follow-up rate of 75\% and an intraclass correlation coefficient for attrition of 0.10. Each plot is the result of 5 replications of the given parameter combination, with a type I error rate of 0.05, two experimental arms, an outcome proportion of 0.30 in the control arm, and 10 clusters of size 20. The vertical lines denote 80\% power. The probabilities of obtaining a power value below 80\%, for each effect size (0.10, 0.20, 0.30, 0.40), are 0.93, 0.54, 0.30, 0.06; the frequentist power estimates (denoted by a point on the x-axis) are 0.21, 0.64, 0.94, 0.99; and the medians of the PFPDs (dotted line) are (0.18, 0.75, 0.97, 0.995) respectively. }
    \label{fig:figure2}
\end{figure*}

\begin{figure*}[ht!]
    \centering
    \includegraphics[width=6in]{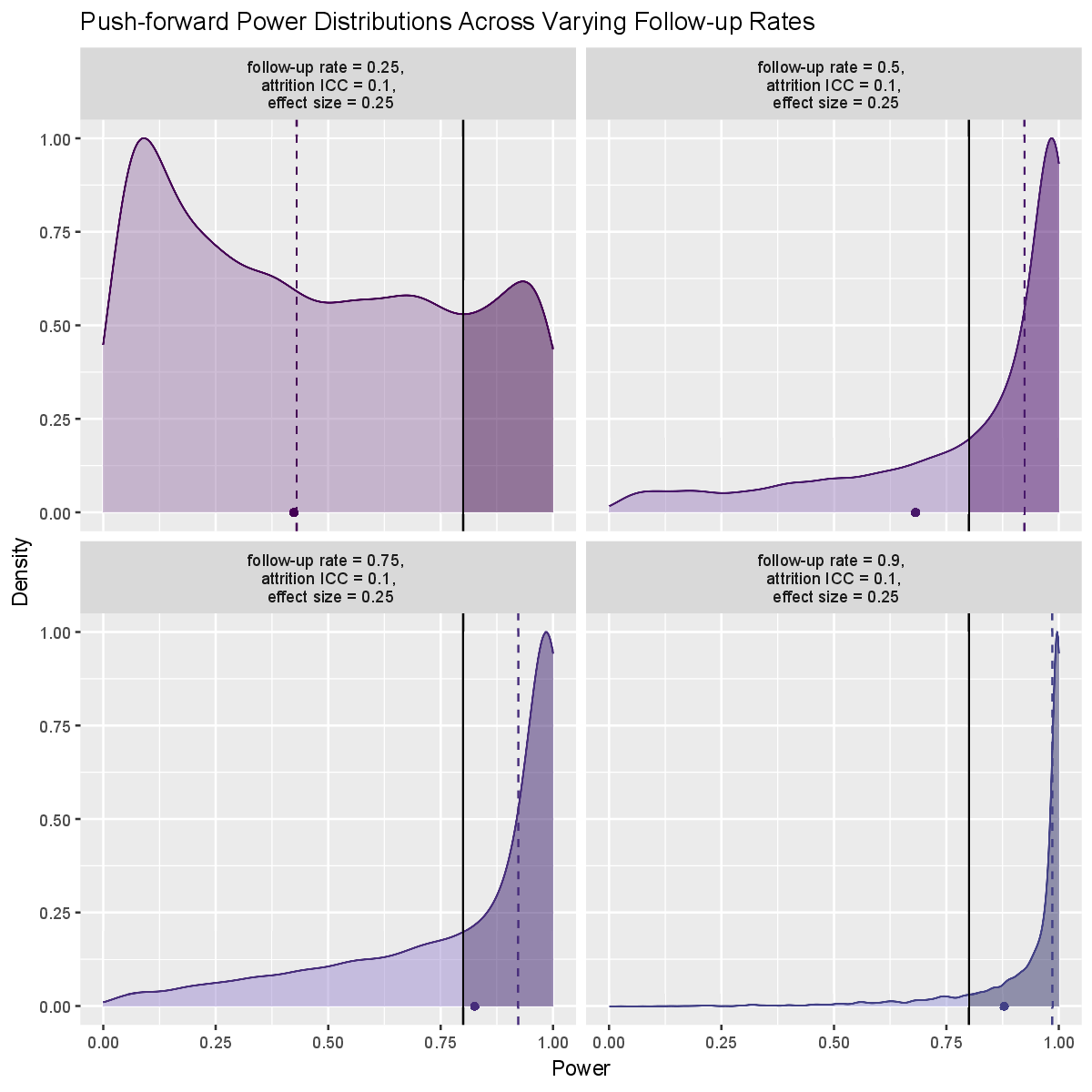}
    \caption{Push forward power distributions across varying follow-up rates given an intraclass correlation coefficient for attrition of 0.10 and a treatment effect of 25\%. Each plot is the result of 5 replications of the given parameter combination, with a type I error rate of 0.05, two experimental arms, an outcome proportion of 0.30 in the control arm, and 10 clusters of size 20. The vertical lines denote 80\% power. The probabilities of obtaining a power value below 80\%, for each follow-up rate (0.25, 0.50, 0.75, 0.90) are 0.80, 0.33, 0.35, 0.11; the frequentist power estimates (denoted by a point on the x-axis) are 0.42, 0.68, 0.82, 0.87; and the medians of the PFPDs (dotted line) are 0.43, 0.92, 0.92, 0.98, respectively. }
    \label{fig:figure3}
\end{figure*}

Figure~\ref{fig:figure2} presents results of the push-forward power approach across varying treatment effect sizes ($\delta_c$), for a given follow-up rate ($\pi_c$) and attrition ICC ($\tau_c$). These results are presented with the frequentist point estimate of power (dot), the median of the PFPD (dashed line), and an 80\% power reference (solid line). Results are similarly presented in Figure~\ref{fig:figure3} for varying follow-up rates. These plots capture information about the PFPD across various possible trial settings (small or medium effect sizes, low or high levels of missingness), and how the frequentist point estimate and PFPD median may relate to the oft used benchmark of 80\% power. 

Summarizing the power as a single value, whether that be the frequentist point estimate or the median of the PFPD, loses important information about the power distribution. For example, the right panel of the middle row of Figure \ref{fig:figure2} indicates that at an effect size of 0.25, the frequentist power estimate of 83\% and the median of the PFPD at 92\% would be deemed sufficient to proceed with the trial. However, there is still a 35\% chance that the power of the trial is below 80\%, which should be taken into consideration when resources are limited. In Figure \ref{fig:figure3}, in the two bottom panels (follow-up rates of 0.75 and 0.90), the frequentist point estimates (0.82 and 0.87) and the PFPD medians (0.92 and 0.98) are fairly similar (within 0.6). However, the probability of obtaining power below 80\% for a follow-up rate of 0.75 is 0.35 and 0.11 for a 0.90 follow-up rate. While the point estimates may seem only marginally improved or even comparable for an increased follow-up rate, there is a 24\% change in the probability of the trial being sufficiently powered according to the PFPD. This may be especially relevant if there is concern around the true follow-up rate in the trial setting.

\clearpage

\section{Application: Design of Bangladesh Study} \label{sec:application}

The development of the push-forward power distribution method is motivated by an upcoming cluster randomized trial in Cox’s Bazar, Bangladesh. The trial will evaluate a community health worker–delivered intervention targeting cardiovascular disease and chronic hepatitis C infection among displaced Rohingya refugees fleeing armed conflict in eastern Myanmar. A related cRCT was recently completed in the conflict-affected region of Karen State, Myanmar, assessing a village health worker-delivered intervention to identify individuals at high risk of cardiovascular disease and deliver care. 

In these and similar settings, attrition is a consistent and complex concern due to unstable and dynamic humanitarian events, such as armed conflict. It is expected that people will move in and out of the study area during the data collection time frame, and entire clusters may be lost to follow-up due to safety concerns or logistical barriers. Alongside these immediate humanitarian threats, chronic and common diseases continue to burden people and create a need for public health interventions. Cardiovascular disease is a leading cause of morbidity and mortality in Myanmar, and attrition in intervention trials further complicates the assessment and implementation of effective care models within a health care system burdened by humanitarian crisis \cite{vos2020, bjertness2016}. We developed the push-forward power distribution approach as a statistical tool for addressing uncertainty and instability when planning trials in humanitarian settings, motivated specifically by the setting of the Bangladesh trial. With a Bayesian framework, we are able to capture uncertainty and nuance that may be particularly interesting to study investigators and stakeholders.

\subsection{Methods and Approach}

Here, we apply our push-forward power distribution methods to the planning of the cRCT in Bangladesh, informed by results from the Myanmar study. Prior to a feasibility study of the Myanmar trial, villages were initially matched into pairs and triplets. For the feasibility study, three triplets were selected from separate clinic areas, and from each triplet a single village was randomly selected to receive the intervention. After establishing feasibility, 4 more villages were randomized to the intervention, and 6 villages were randomized to the control arm for a total of 13 villages participating in the study. Participants in the intervention arm were assessed for eligibility to participate in the longitudinal study. Those individuals already enrolled in the feasibility study were invited to re-enroll in the longitudinal cRCT. The intervention consists of an initial screening to assess prior history of myocardial infarction, stroke, diabetes, hypertension, or elevated cardiovascular disease risk. Eligible participants were then recommended for confirmatory and follow-up visits to treat and monitor their CVD-related health conditions. In the control arm, data was collected via cross-sectional surveys. The primary outcome was adherence to medication (blood pressure management medication and/or statin). Further details about the Myanmar study may be found in the associated manuscript \cite{thwe2026}. 

\begin{figure}[ht!]
    \centering
    \includegraphics[width=0.5\linewidth]{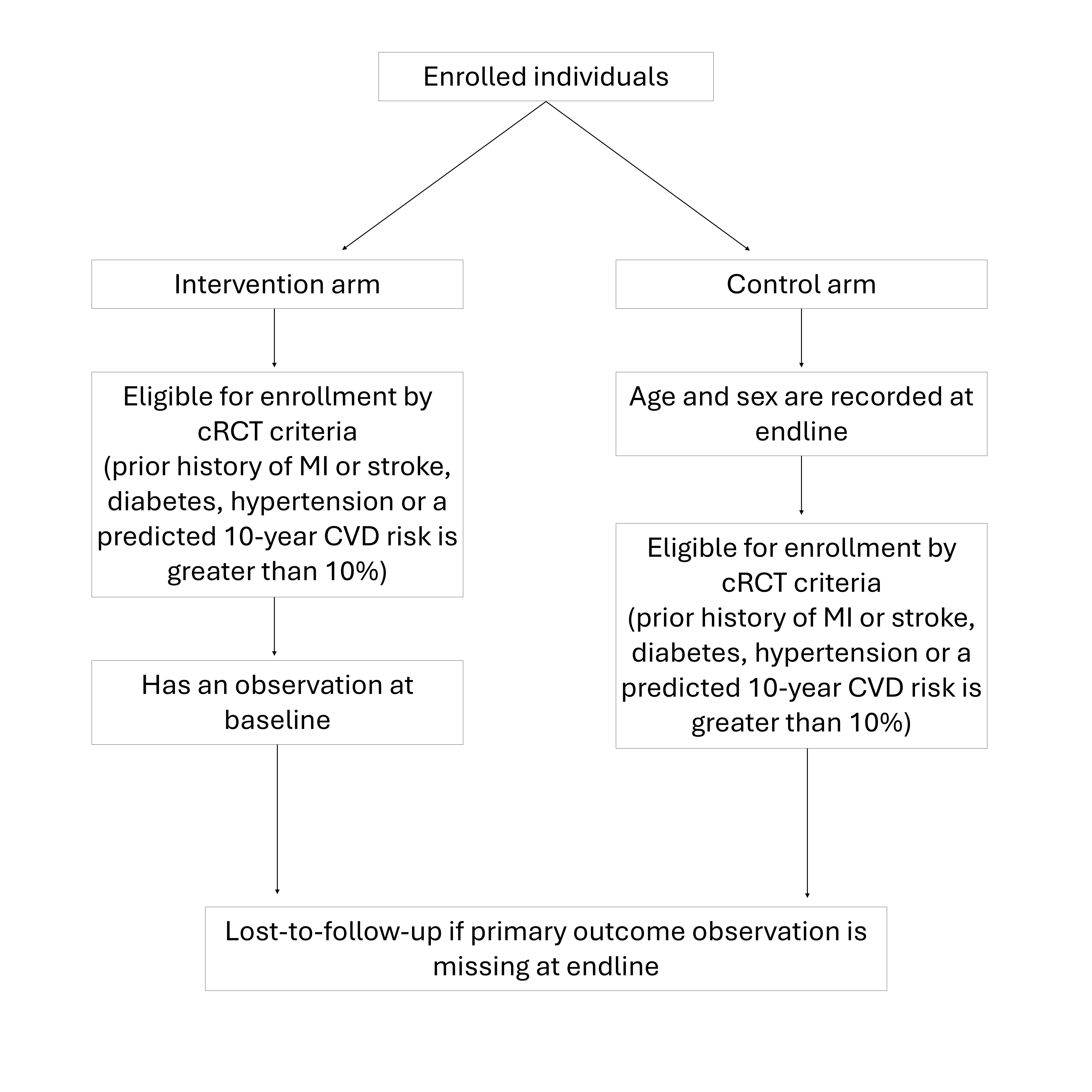}
    \caption{Flow diagram with details of how data from the Myanmar study was selected for inclusion in our application. We included individuals who met the eligibility criteria for the cRCT and were expected to have an observation at endline.}
    \label{fig:figure4}
\end{figure}

We analyze data from the Myanmar trial to estimate follow-up rate, attrition ICC, and treatment effect on adherence to medication. We included data from individuals in the intervention arm if they were eligible for enrollment in the longitudinal cRCT and had a primary outcome observation at baseline. Individuals from the control arm were included if their age and sex were observed, and they met the criteria for the cRCT (Figure \ref{fig:figure4}). The final data set includes 310 individuals (223 in the intervention arm and 87 in the control arm). In the intervention arm, 34 individuals (15\%) were missing a primary outcome due to death, moving out of the study area, or other causes. Of note, 12 of these individuals lived in a village that was entirely lost to follow-up due to conflict activity. In the control arm, there was no simple approach to estimating level of missingness due to the cross-sectional design. 

To better inform the push-forward power results for the Bangladesh trial, we considered various reasonable scenarios by imposing different follow-up rates and attrition correlation structures on the control arm, as the study was not able to easily assess missingness in that arm. The scenarios are as follows:

\begin{enumerate}
  \item[] \textbf{Scenario 1}: 10\% of individuals in the control arm are missing at random
  \item[] \textbf{Scenario 2}: 15\% of individuals in the control arm are missing at random
  \item[] \textbf{Scenario 3}: 20\% of individuals in the control arm are missing at random
  \item[] \textbf{Scenario 4}: 10\% of individuals in the control arm are missing due to a cluster dropping out, the rest are missing at random
  \item[] \textbf{Scenario 5}: 15\% of individuals in the control arm are missing due to multiple clusters dropping out
  \item[] \textbf{Scenario 6}: 20\% of individuals in the control arm are missing due to multiple clusters dropping out
\end{enumerate}

\begin{table}[ht!]
    \centering
    \begin{tabular}{c|c|c|c|c}
        Arm & Village ID & No. Missing & No. Observed & Total \\\hline
        Control & 1 & 0 & 16 & 16 \\
                & 2 & 0 & 17 & 17 \\
                & 3 & 0 & 14 & 14 \\
                & 4 & 0 & 19 & 19 \\
                & 5 & 0 & 17 & 17 \\
                & 6 & 0 & 4 & 4 \\\hline
                & Total & 0 & 87 & 87 \\\hline
   Intervention & 7 & 1 & 22 & 7 \\
                & 8 & 5 & 29 & 15 \\
                & 9 & 3 & 12 & 6 \\
                & 10 & 1 & 7 & 31 \\
                & 11 & 9 & 42 & 28 \\
                & 12 & 4 & 44 & 19 \\
                & 13 & 11 & 33 & 44 \\\hline
                & Total & 34 & 189 & 223 \\\hline 
    \end{tabular}
    \caption{Summary of attrition data from the Myanmar study used to inform power calculations for the Bangladesh trial, stratified by village and experimental arm.}
    \label{tab:attrition-data}
\end{table}

\clearpage

\subsection{Application Results}

\begin{figure}[ht!]
    \centering
    \includegraphics[width=0.90\linewidth]{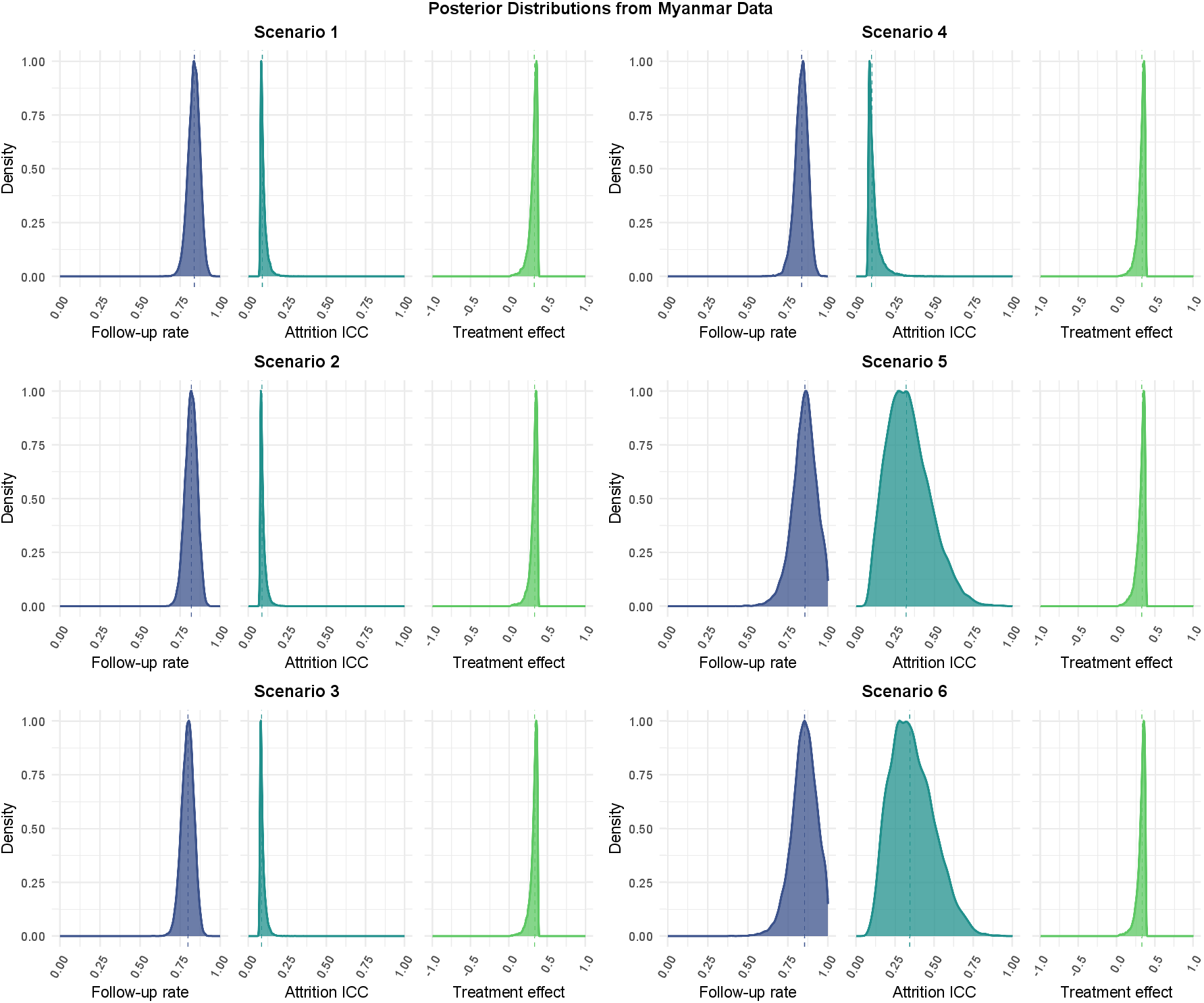}
    \caption{Posterior distributions estimated using the Myanmar study data for each of the six control arm missingness structures. For scenarios 1-6, the median follow-up rate (indicated by a dashed vertical line) and 95\% CRs are 0.84 [0.76, 0.90], 0.82 [0.74, 0.89], 0.80 [0.72, 0.87], 0.84 [0.75, 0.91], 0.86 [0.69, 0.98], and 0.85 [0.68, 0.98]. The median attrition ICC and 95\% CRs are 0.09 [0.08, 0.15], 0.09 [0.07, 0.14], 0.09 [0.07, 0.14], 0.10 [0.08, 0.23], 0.32 [0.12, 0.64], and 0.34 [0.14, 0.66]. The median treatment effect sizes and 95\% CRs are 0.33 [0.17, 0.38], 0.34 [0.18, 0.37], 0.34 [0.19, 0.38], 0.33 [0.18, 0.37], 0.33 [0.16, 0.37], and 0.33 [0.17, 0.37].}
    \label{fig:figure5}
\end{figure}

\begin{figure}[ht!]
    \centering
    \includegraphics[width=0.75\linewidth]{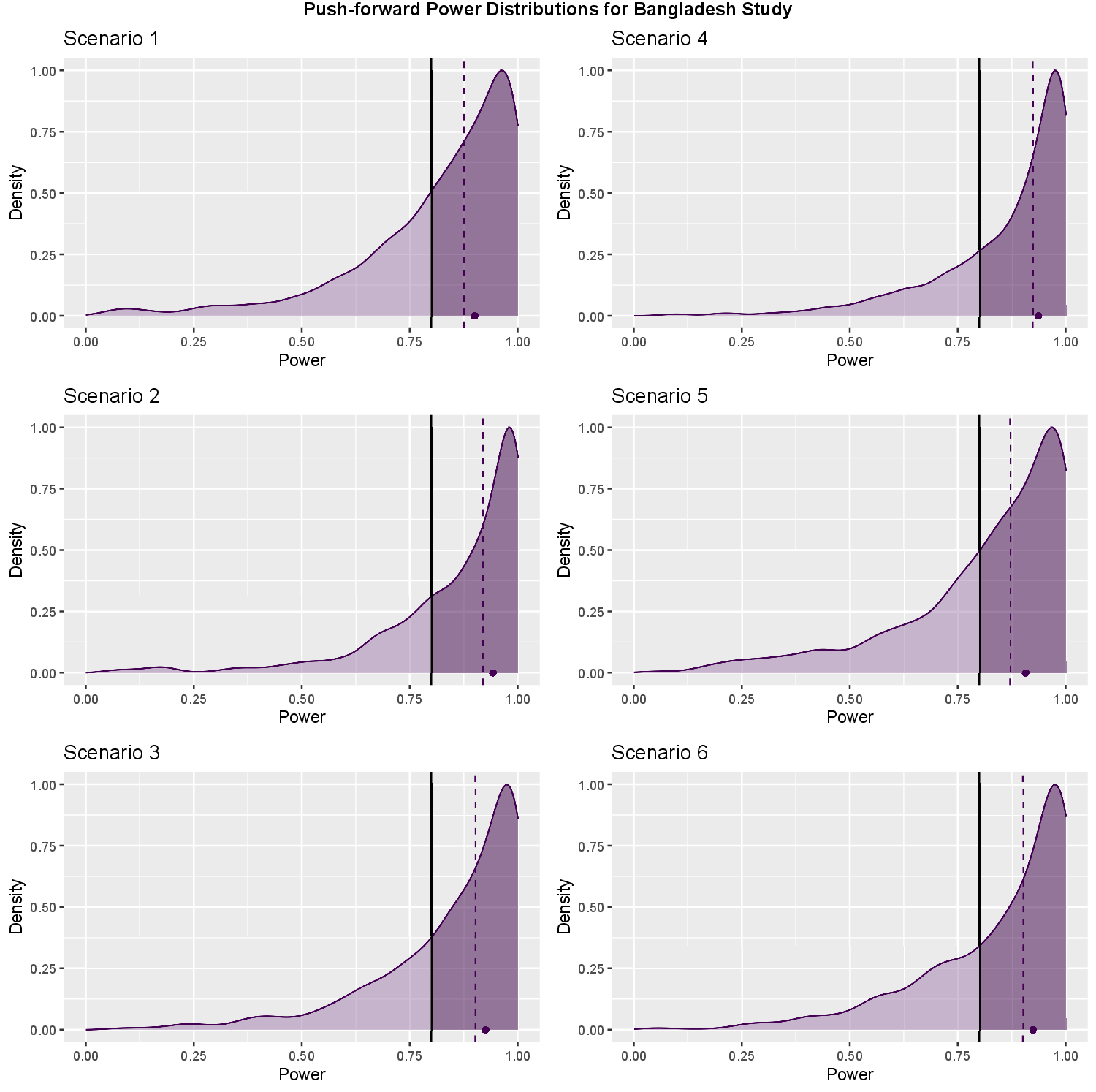}
    \caption{Push forward power distribution for the planned CVD study in Bangladesh, informed by data from a similar trial in Myanmar. The solid vertical line indicates 80\% power, the dashed line denotes the median of each PFPD, and the point on the x-axis denotes the frequentist power estimate. The probabilities of achieving power below 80\% with 24 clusters of size 15 for each of the six scenarios are 0.34, 0.25, 0.29, 0.25, 0.35, and 0.31. The median power for each of the scenarios is 0.87, 0.92, 0.90, 0.92, 0.87, and 0.90. }
    \label{fig:figure6}
\end{figure}

For each scenario, we proceeded with Bayesian estimation of the key parameters (Figure \ref{fig:figure5}) and construction of the push-forward power distribution (Figure \ref{fig:figure6}). The median estimated follow-up rates range from 0.80 to 0.86, depending on the imposed attrition structure. For scenarios with full cluster drop-out (5 and 6), we see more uncertainty and larger credible regions of the follow-up rate posterior distributions. As we would expect, the attrition ICCs are very low for scenarios 1-4, where attrition is fully or partially at random and not correlated with the clustering structure. For scenarios 5 and 6, the posterior distributions for the ICC are shifted closer to 1 and are much wider, reflecting more uncertainty and more within-cluster attrition correlation. For all scenarios, the posteriors of the treatment effect size remain fairly similar and consistent, centered around 0.33. 

The push-forward power distributions show reasonable expected power across the various attrition scenarios with 12 clusters per arm and 15 individuals per cluster. We can use the PFPDs to make probabilistic claims about the planned power for the trial, i.e. the highest probability of achieving power below 80\% is 0.35, for scenario 5 (Figure \ref{fig:figure6}). The medians of the PFPDs are all at or above 0.87, with most of the density concentrated closely around the median and above 0.80. These positive and consistent results are driven by the large treatment effect size which is seen across all scenarios, despite fluctuations in the attrition parameter posteriors. We may interpret these results in comparison to the simulation PFPD results in Figure \ref{fig:figure2}, where we see much more uncertainty and fluctuation in the power distributions. For the Bangladesh study, even across various possible attrition scenarios, we may be reasonably confident that the trial will achieve sufficient power, as the PFPDs appear stable with a high probability of achieving power above 80\%.  

\clearpage

\section{Discussion} \label{sec:discussion}

In the field of humanitarian experimental studies, there is a need for more rigorous statistical and quantitative trial methods focused on complex settings. Here, we propose a Bayesian approach to power calculations for cRCTs in humanitarian settings designed to explicitly incorporate uncertainty in key parameter estimates and prior information from existing sources. We provide our methods as a Shiny app in R, to promote accessibility and usability for practitioners. The self-contained Shiny app and all other code associated with this work is available at \url{https://github.com/melliott432/cRCT_PFPD}.  This work is motivated by an upcoming trial in Cox's Bazaar, Bangladesh, where attrition of clusters is a key concern. We apply our methods to the planning of this trial using data from a similar completed trial in Myanmar. We also investigate our methods with a simulation study and compare our results with a typical frequentist approach to power calculations. 

Our sample size simulation study demonstrates how uncertainty in power changes with sample size, and how this information may be more fully captured with a distributional representation using the Bayesian paradigm. Across a varying number of clusters, cluster size, and attrition ICC, changes in the shape and peaks of the PFPD display important information about planned power and provide more nuance when comparing sample size options. For example, there is an implied diminishing benefit to increasing cluster size after a certain total sample size is reached; i.e. for 10 clusters, we see a large change in the PFPD as the cluster size increases from 25 to 100, but a negligible visual difference when increasing from a cluster size of 100 to 200. 

Importantly, the PFPDs presented in the simulation study highlight a critical limitation of traditional frequentist power calculations. Even when the average of the power distribution exceeds a conventional threshold (such as 80\%), the tail of the distribution may contain a non-negligible probability, and the probability of obtaining power below 80\% may be unacceptably high. Therefore, the distributions provide insights on the risks that the actualized trial power will fall below acceptable limits. In complex settings, understanding the risk of under-powering a trial is essential for study design. The PFPD provides a readily-understandable visual to aid conversations about power between investigators and stakeholders. Investigators may interactively explore the PFPD behavior across a wider range of sample sizes and parameter configurations via the R Shiny application.

In our application to the Bangladesh trial, we see that our push-forward method aligns with a frequentist approach on measures of central tendency. However, the push-forward distribution provides a fuller picture of uncertainty due to expected attrition, which may be unpredictably exacerbated by crisis settings. Therefore, our methods can contribute to a more thorough and nuanced approach to planning cluster randomized trials in humanitarian settings by quantifying uncertainty and prior information about key concerns such as attrition. The plots presented help visualize how power changes with other operating characteristics, and could be used alongside point estimates to guide decisions when planning a trial. Investigators may examine with a probabilistic mindset how much power is lost or gained when changing sample size. 

Limitations to our work include the need for further investigation of our methods across different outcome types and models and an application with multiple sources of historical data. We present results for a dichotomous outcome here, but similar investigation of our methods for a continuous outcome is needed. For our simulations and real-world application, we use only a single prior information source due to limited availability of entire data sets from relevant settings. Our results here are limited by a lack of demonstration of the proposed hierarchical modeling structure. Additionally, our application analysis assumes the data are missing completely at random due to limited data; however, the proposed approaches can be easily extended to other missingness settings beyond those presented.

In humanitarian crisis settings, it is important to account for expected but unpredictable events that may impact the power of a planned trial through attrition or other related effects. With the results presented here, we demonstrate that quantifiable uncertainty about planned power may be lost when only a frequentist point value is used. Using a distributional Bayesian representation, such as a push-forward power distribution, can also provide important additional information in a visually concise format.  With the methods proposed here, we hope to contribute to more rigorous power approaches that account for uncertainty in a structured statistical format and address a critical gap highlighted by humanitarian practitioners. 

\section{Acknowledgements}
Funding for the Myanmar and Bangladesh trials is supported by Research for Health in Humanitarian Crises (R2HC), Elrha \#78698; Novo Nordisk Foundation NNF24OC0091448; and National Institutes of Health grants: R01 AI-147441, R01 AI-13890, and P30 AI-036214. The authors would also like to thank trial participants and stakeholder workshop participants.

\newpage

\onecolumn
 
\section*{Appendix}

\setcounter{table}{0}
\renewcommand{\thetable}{A\arabic{table}}

\subsection*{A.1}

\textbf{Study design summary table}

\begin{table}[ht!]
    \centering
    \begin{sideways}
    \scriptsize
    \setlength{\tabcolsep}{4pt}
    \begin{tabular} {|L{0.9in}|L{1.0in}|L{1.2in}|L{0.85in}|L{1.55in}|L{1.2in}|L{1.45in}|}\hline
    \textbf{Study} 
    & \textbf{Health Setting}	
    & \textbf{Crisis Setting} 
    & \textbf{Design Type} 
    & \textbf{Design Considerations} 
    &	\textbf{Pre-Specified Adaptations} 
    &	\textbf{Ad hoc Adaptations} 
    \\\hline
        Walton et al. 2025 
    & Nutrition, child wasting 
    & Climate crisis, food insecurity, economic stress in Somalia 
    & Adaptive cRCT 
    & —
    & Adjusted timeline; Reduced no. of clusters due to flooding, eligibility criteria
    & Adapting instruments; Flexible data collection during flooding 
\\\hline
          Fenn et al. 2017
    & Child and maternal nutrition
    & Flooding, drought, conflict, food insecurity in Pakistan
    & Longitudinal cRCT
    & —
    & —
    & 27 enrolled households migrated and were not replaced; Contextual difficulties for data collection (heat, personnel)
    \\\hline
          Leroy et al. 2021
    & Health and nutrition, child wasting
    & Poverty, history of genocide and political unrest in Burundi
    & cRCT with cross-sectional data
    & —
    & —
    & — 
    \\\hline
          Jordans et al. 2020
    & Child and maternal nutrition, child wasting
    & Rural flooding and civil war in Nepal
    & cRCT
    & Buffer zone for contamination (due to nature of intervention)
    & —
    & —
    \\\hline
          Kurdi et al. 2020
    & Mental health
    & Civil conflict in Yemen
    & cRCT
    & Number of clusters informed by low expected uptake (due to social, financial)
    & —
    & 95 replacement households added, follow-up delayed due to conflict
    \\\hline
          Watson et al. 2023
    & Malnutrition
    & IDP camps in Somalia
    & cRCT
    & Covariate constrained randomization used to balance 3 covariates between clusters, including binary disaster risk
    & —
    & —
    \\\hline
          Tucker et al. 2024
    & Mental health
    & War in Ukraine
    & Pragmatic cRCT
    & —
    & Control group delivery timing due to conflict
    & —
    \\\hline
          Rahman et al. 2016
    & Mental health
    & Armed conflict in Pakistan
    & cRCT
    & Proportionate stratified sampling by IDP camp
    & —
    & —
    \\\hline
          Jordans et al. 2010
    & Mental health
    & Poverty and post-war political instability in rural Nepal
    & cRCT
    & Sample size accounted for attrition due to active war.
    & —
    & —
    \\\hline
          Tol et al. 2014
    & Mental health in children
    & Armed conflict in Burundi
    & cRCT
    & Flexible intervention (timing and format) to accommodate conflict and movement
    & — 
    & —
    \\\hline
          Kim et al. 2023
    & Mental health in children
    & Refugees from Syrian civil war living in Lebanon
    & cRCT
    & Low follow-up rates (35\%,44\%) due to security threats; MI sensitivity analysis for missing data
    & — 
    & —
    \\\hline
          Grijalva-Eternod et al. 2023
    & Psychosocial
    & IDP camps in Somalia
    & Non-randomised controlled cluster trial 
    & 
    & —
    & More clusters included due to less children being present than expected
    \\\hline
          Roberts et al. 2001
    & Malnutrition
    & Refugee camp in Malawi
    & cRCT
    & Provinces selected based on continued conflict, homogeneous population, existing trained personnel; excluded areas with active conflict
    & — 
    & —
    \\\hline
    \end{tabular}
    \end{sideways}
    \caption{Summary of experimental studies of health interventions in humanitarian settings, including adaptations to trial design. IDP = internally displaced people, cRCT = cluster-randomized controlled trial}
    \label{tab:tableA1}
\end{table}

\clearpage

\subsection*{A.2}

\textbf{Reparameterization of $\tau$ for Beta distribution}

\begin{align}
\tau &= \frac{Var(R^{t}_j)}{Var(R^{t}_{ij})} \\
\notag &= \frac{\alpha \beta}{(\alpha + \beta)^2(\alpha + \beta + 1)} \times \bigg[\frac{\alpha}{\alpha + \beta}\bigg(\frac{\alpha + \beta - \alpha}{\alpha + \beta}\bigg)\bigg]^{-1}\\
\notag &= \frac{\alpha \beta}{(\alpha + \beta)^2(\alpha + \beta + 1)} \bigg[\frac{(\alpha + \beta)^2}{\alpha \beta}\bigg]\\ 
\notag &= \frac{1}{\alpha + \beta + 1}
\end{align}

\subsection*{A.3}
\textbf{Shiny App}

\begin{figure}[ht!]
    \centering
    \includegraphics[width=1\linewidth]{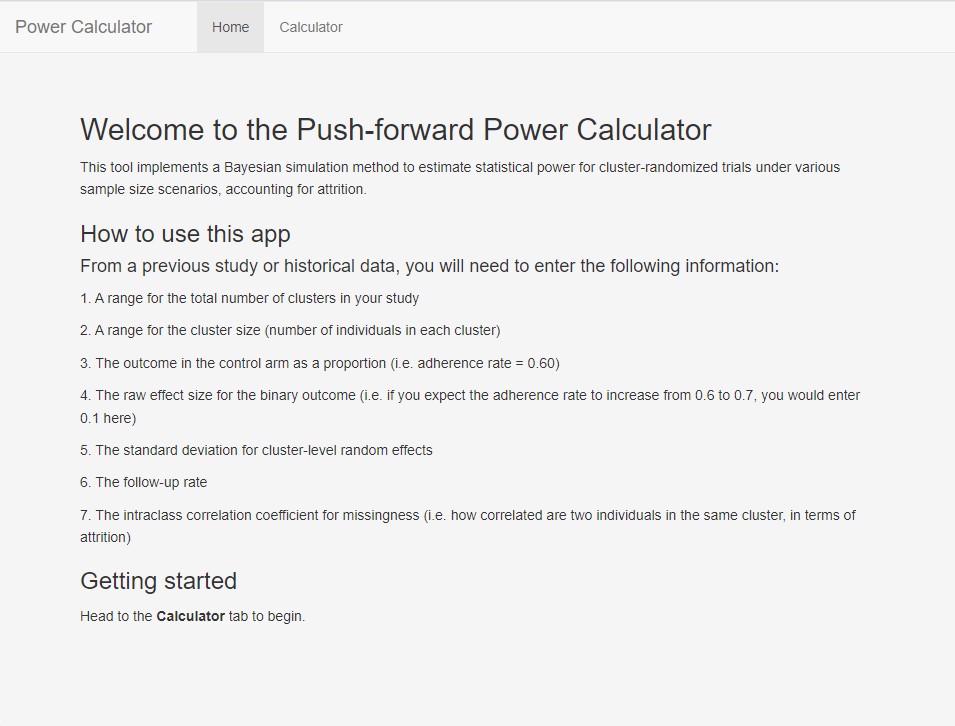}
    \caption{Landing page of the R Shiny app.}
    \label{fig:a31}
\end{figure}

\begin{figure}[ht!]
    \centering
    \includegraphics[width=1\linewidth]{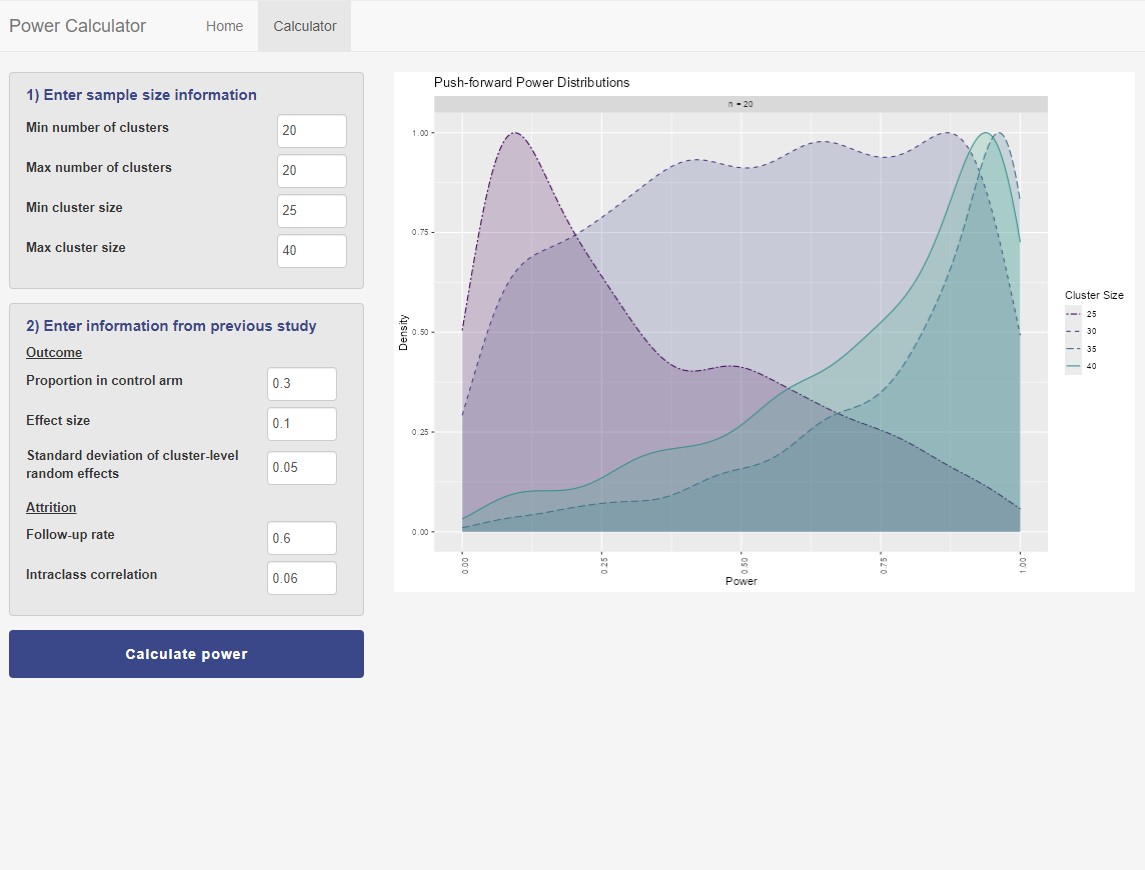}
    \caption{Calculator page of the R Shiny app, showing app output as a plot. }
    \label{fig:a32}
\end{figure}

\end{document}